\documentstyle[12pt]{article} 
\def\be{\begin{eqnarray}} \def\ee{\end{eqnarray}} \topmargin -0.3in
\begin{document} 
\pagestyle{empty}
\Huge{\noindent{Istituto\\Nazionale\\Fisica\\Nucleare}}

\vspace{-3.9cm}

\Large{\rightline{Sezione SANIT\`{A}}} \normalsize{} \rightline{Istituto
Superiore di Sanit\`{a}} \rightline{Viale Regina Elena 299} \rightline{I-00161
Roma, Italy}

\vspace{0.65cm}

\rightline{INFN-ISS 97/11} \rightline{October 1997}

\vspace{0.5cm} \large

\begin{center}{\large \bf Deep inelastic scattering and final state interaction
in an exactly solvable relativistic model }

\vspace{.5cm}

E.  Pace$^1$, G.  Salm\`e$^2$, F.M.  Lev$^3$

{\small \it $^1$Dipartimento di Fisica, Universit\`a di Roma "Tor Vergata" and
INFN, Sezione Tor Vergata, Via della Ricerca Scientifica 1, I-00133, Rome,
Italy}

{\small \it $^2$INFN, Sezione Sanit\`a, Viale Regina Elena 299, I-00161 Rome,
Italy}

{\small \it $^3$ Laboratory of Nuclear Problems, Joint Institute for Nuclear
Research, Dubna, Moscow region 141980, Russia} \end{center}

\begin{abstract}

In the theory of deep inelastic scattering (DIS) the final state interaction
(FSI) between the struck quark and the remnants of the target is usually assumed
to be negligible in the Bjorken limit.  This assumption, still awaiting a full
validation within nonperturbative QCD, is investigated in a model composed by
two relativistic particles, interacting via a relativistic harmonic oscillator
potential, within light-cone hamiltonian dynamics.  An electromagnetic current
operator whose matrix elements behave properly under Poincar\'e transformations
is adopted.  It is shown that:  i) the parton model is recovered, once the
standard parton model assumptions are adopted; and ii) when relativistic,
interacting eigenfunctions are exactly taken into account for both the initial
and final states, the values of the structure functions, averaged over small,
but finite intervals of the Bjorken variable $x$, coincide with the results of
the parton model in the Bjorken limit.

\begin{flushleft}{\bf PACS}:  12.39.Pn, 13.60.Hb \end{flushleft}
\begin{flushleft} {\it Key words:  deep inelastic scattering, confinement,
relativistic harmonic oscillator.}  \end{flushleft} \end{abstract} 
\vspace{2cm}
\hrule width5cm \vspace{.2cm} \noindent{\footnotesize{ {\bf Phys.  Rev.  C} in
press.}}

\newpage \normalsize \pagestyle{plain}
\section{Introduction} \label{S1}

 The parton model proposed by Bjorken and Feynman almost thirty years ago
\cite{Bjor,Feyn} has turned out to be a convenient language for discussing
different properties of deep inelastic scattering (DIS), although, according to
the theory based on the operator product expansion \cite{Wil}, the parton model,
even in the Bjorken limit, is accurate only up to anomalous dimensions and
perturbative QCD corrections.

If $q$ is the momentum transfer in DIS, $Q=|q^2|^{1/2}$, $P'$ is the
four-momentum of the target in the initial state and $m$ is its mass, then the
Bjorken limit is defined by the condition that $Q\gg m$ and the Bjorken variable
$x=Q^2/(2 P'q)$ is not too close to 0 and 1.  In the parton model, it is assumed
that the final state interaction (FSI) of the struck quark with the remnants of
the target is a higher twist effect, i.e., an effect which is suppressed at
least as $m^2/Q^2$.  The qualitative motivation of this assumption is that the
time needed for the absorption of the virtual photon by the struck quark is much
smaller than the time of its hadronization and therefore in the process of
absorption the struck quark can be considered as approximately free.  It has
been also shown within the framework of the collinear expansion \cite{EFP} that
the Feynman diagrams describing the FSI are indeed suppressed at high $Q^2$,
once proper assumptions are adopted.

It could be thought very surprising that FSI can be disregarded in presence of
confinement (see, e.g., Ref.  \cite{Gurv}).  Indeed, at first glance the picture
which follows from confinement fully differs from that given by the parton
model, i.e., no interaction in the final state and a continuous spectrum.  In
particular, while the structure functions in the parton model are continuous, in
models where confinement is taken into account the structure functions are
linear combinations of delta functions.  However the two models can be
reconciled within the framework of the theory of distributions.  An analogous
situation takes place in the case of the reaction $e^+e^-\rightarrow hadrons$
(see, e.g., the discussion in Ref.  \cite{sim} and references therein).

 Of course the relevance of the FSI has to be studied in the framework of
nonperturbative QCD, but in absence of a full solution it is desirable to
consider models in which the structure functions can be calculated exactly and
therefore it is possible to check whether the FSI is indeed a higher twist
effect.

 In the last years several nonrelativistic two-body models of inelastic
scattering were considered (see, e.g., Refs.  \cite{IKL,PSW}) and it was shown
that the effects of the FSI in these models are in agreement with the standard
parton model assumptions, even in the case of confining interactions.  However,
DIS in the Bjorken limit by no means can be considered nonrelativistically.
Indeed, in this limit the energy transfer $q^0$ in the target rest frame has to
satisfy the condition $q^0\gg m$, while the nonrelativistic approach holds only
if $q^0\ll m$.

 The FSI for a relativistic particle bound in an external field was considered
in Ref.  \cite{IKL}.  However, in this case one has a two-body problem, where
one of the particles (the source of the field) has an infinitely large mass.
Therefore this problem cannot be studied in the Bjorken limit, where $Q$ is much
larger than all the masses involved in the problem.  Another relativistic
approach to the FSI has been considered in Ref.  \cite{Gurv} in the framework of
the Bethe-Salpeter formalism, but it was assumed, as usual, that only Feynman
diagrams from a certain class (so called handbag diagrams) are dominant.  Then
the result of Ref.  \cite{EFP} that the FSI is a higher twist effect has been
confirmed.

 The aim of this paper is to investigate the role played by FSI in DIS for an
exactly solvable relativistic constituent quark model (CQM), within the
light-cone (= front-form) hamiltonian dynamics.  As well known, in the framework
of relativistic CQMs (see, e.g., Ref.  \cite{CI}) confinement can be ensured by
choosing a quark-quark potential such that the mass operator of a system with a
fixed number of relativistic constituent quarks has only the discrete spectrum
(while in QCD confinement is understood as the property of the quark and gluon
Green functions to have no poles for real values of the mass).  Our purpose is
to verify whether the naive treatment of confinement in relativistic CQMs is
compatible with the parton model.

 We consider a simple system composed by two relativistic particles interacting
via the relativistic harmonic oscillator potential.  We adopt an electromagnetic
current whose matrix elements exhibit the correct properties under Poincar\'e
transformations and fulfill the current conservation, as shown in \cite{LPS}.
In the proper Breit frame, the relevant components of the current are the same
as in the parton model.  Then, in the framework of the light-cone hamiltonian
dynamics, we can derive exact expressions for the DIS structure functions,
including the FSI effects, and show that in the Bjorken limit the exact results
coincide with those given by the parton model, after an average over small
intervals of the scaling variable $x$ has been performed.  This average features
the finite detector resolution and allows us to avoid some mathematical
technicalities of the theory of distributions.

 The paper is organized as follows.  In Sec.  \ref{S2} we explicitly define a
model describing the interaction of two relativistic spin $1/2$ particles, then
in Sec.  \ref{S3} we calculate the DIS structure functions of this system with
the standard parton model assumptions and in Sec.  \ref{S4} the structure
functions are calculated using the exact two-body wave functions.  Conclusions
are drawn in Sec.  5.

\section{Relativistic harmonic oscillator potential} \label{S2}

 We consider a system of two different particles with the same mass, $m_0$, and
spin $1/2$.  To describe such a system it is necessary to choose first an
explicit form of the unitary irreducible representation (UIR) of the Poincar\'e
group pertaining to each particle.  There are many equivalent ways to construct
an explicit realization of such a representation \cite{Wig}.  We choose the
realization in the front form of dynamics (see, e.g., Refs.  \cite
{BarHal,Ter,CP,KP,Lev}).

  Let $p$ be the particle 4-momentum, ${\vec s}$ be the spin operator and
$\sigma$ be the spin projection on the $z$ axis ($\sigma =\pm 1/2$).  We define
$p^{\pm}=(p^0\pm p^z)/ \sqrt{2}$, and we use ${\vec p}_{\bot}$ to denote the
projection of $p$ onto the plane $xy$.  The one particle Hilbert space can be
chosen as the space of functions $\phi({\vec p}_{\bot},p^+,\sigma)$ such that

\begin{eqnarray} &&(\phi,\phi)=\sum_{\sigma}\int\nolimits |\phi({\vec
p}_{\bot},p^+,\sigma)|^2d\rho({\vec p}_{\bot},p^+)<\infty,\nonumber\\
&&d\rho({\vec p}_{\bot},p^+)=\frac{d{\vec p}_{\bot}dp^+}{2(2\pi)^3p^+} \label{1}
\end{eqnarray}

 The Hilbert space $\cal H$ for the representation of the Poincar\'e group
describing a system of two free or interacting particles is realized in the
space of functions $ \phi ({\vec p}_{1\bot},p_1^+, \sigma_1, {\vec
p}_{2\bot} ,p_2^+, \sigma_2)$ such that

\begin{equation} \sum_{\sigma_1 \sigma_2}\int\nolimits |\phi({\vec
p}_{1\bot},p_1^+,\sigma_1,{\vec p}_{2\bot},p_2^+,\sigma_2)|^2 \prod_{i=1}^{2}
d\rho({\vec p}_{i\bot},p_i^+) \, <\, \infty \label{3} \end{equation}

  Instead of the variables ${\vec p}_{1\bot}$, $p_1^+$, ${\vec p}_{2\bot}$,
$p_2^+$, let us consider the variables ${\vec P}_{\bot}$, $P^+$, and ${\vec k}$,
where ${\vec P}_{\bot}={\vec p}_{1\bot}+ {\vec p}_{2\bot}$, $P^+=p_1^++p_2^+$
and the relative momentum ${\vec k}$ is defined as follows \cite {Ter,KP}.  We
first define the quantities

\begin{equation} \xi=\frac{p_1^+}{P^+},\quad {\vec k}_{\bot}={\vec p}_{1\bot}-
\xi {\vec P}_{\bot} \label{4} \end{equation}

\noindent and then define ${\vec k}=({\vec k}_{\bot},k_z)$, where

\begin{equation} \xi -\frac{1}{2}=\frac{k_z}{2\omega({\vec k})},\quad
\omega({\vec k})=(m_0^2+{\vec k}^2)^{1/2}.  \label{5} \end{equation}

A direct calculation shows that

\begin{eqnarray} &&\prod_{i=1}^{2}d\rho_i({\vec p}_{i\bot},p_i^+)= d\rho({\vec
P}_{\bot},P^+)d\rho(int),\nonumber\\ &&d\rho(int)=\frac{d{\vec
k}_{\bot}d\xi}{2(2\pi)^3\xi (1-\xi)}= \frac{d{\vec k}}{(2\pi)^3 \omega({\vec
k})} \label{6} \end{eqnarray}

\begin{sloppypar} If the particles do not interact with each other, the
 generators of the two-particle representation are equal to sums of the
 corresponding one-particle generators \cite{Ter}.  The result is that the
 free-mass operator of the system is $M_0=2\omega({\vec k})$ and the two-body
 spin operator is equal to

\begin{equation} {\vec S}=U^{-1}[{\vec l}({\vec k})+{\vec s}_1+{\vec
s}_2]U,\quad U=v({\vec k},{\vec s}_1)v(-{\vec k},{\vec s}_2), \label{8}
\end{equation}

\noindent where $v({\vec k},{\vec s})$ is the Melosh matrix \cite{Mel}.  In the
given context the Melosh matrix was first considered by Terent'ev \cite{Ter}:

\begin{equation} v({\vec k},{\vec s})=exp\bigl(\frac{2\imath
\epsilon_{jl}s^jk^l} {k_{\bot}}~arctg\frac{k_{\bot}}{m_0+\omega({\vec
k})+k_z}\bigr) \label{9} \end{equation}

\noindent with $k_{\bot}=|{\vec k}_{\bot}|$, $j=1,2$, $l=1,2$, and
$\epsilon_{12}=-\epsilon_{21}=1,\epsilon_{11}=\epsilon_{22}=0$.  \end{sloppypar}

  Let us define the "internal" Hilbert space ${\cal H}_{int}$ as the space of
functions $\chi=\chi({\vec k},\sigma_1,\sigma_2)$ such that

\begin{equation} \sum_{\sigma_1\sigma_2} \int\nolimits |\chi({\vec
k},\sigma_1,\sigma_2)|^2 d\rho(int)\quad < \infty \label{10} \end{equation}

 If particles 1 and 2 interact with each other, then the representation space is
the same, but the generators differ from the free ones (see, e.g., Eqs.
(A34-A36) of Ref.  \cite {CP} or Eq.  (2.34) of Ref.  \cite {Lev}).  The
interaction should be introduced in such a way that the new set of generators,
as well as the set of free generators, satisfies the commutation relations of
the Poincar\'e group Lie algebra.  By analogy with Ref.  \cite{BT}, this can be
done by replacing the free mass operator $M_0$ by an interaction dependent mass
operator $M$, which acts only through ${\vec k}, \sigma_1, \sigma_2$ and
commutes with the spin operator ${\vec S}$ given by Eq.  (\ref{8}).  As well
known, the system of generators obtained in such a way defines the front form of
dynamics \cite{Dir}, in which only three generators are interaction dependent
and the other seven generators are free.

 Then, let us define the internal space ${\cal H}_{int}'$, whose elements
$\Psi({\vec k},\sigma_1,\sigma_2) = \langle {\vec
k},\sigma_1,\sigma_2|\Psi\rangle$ are normalized as in the nonrelativistic
quantum mechanics:

\begin{equation} \sum_{\sigma_1\sigma_2} \int\nolimits |\Psi({\vec
k},\sigma_1,\sigma_2)|^2 \frac{d^3{\vec k}}{(2\pi)^3}\quad < \infty \label{11}
\end{equation}

\noindent As follows from Eqs.  (\ref{6}), (\ref{8}), (\ref{10}) and (\ref{11}),
if the relation between the spaces ${\cal H}_{int}$ and ${\cal H}_{int}'$ is
defined as \begin{equation} \chi({\vec k},\sigma_1,\sigma_2) = \langle {\vec
k},\sigma_1,\sigma_2|\chi\rangle = \langle {\vec k},\sigma_1,\sigma_2|U^{-1}
|\Psi\rangle \omega({\vec k})^{1/2}, \label{12} \end{equation}

\noindent then the two-body spin operator in ${\cal H}_{int}'$ has the standard
"non-relativistic" form ${\vec S}_{nr}={\vec l}({\vec k})+{\vec s}_1+{\vec s}_2$
(that also coincides with the instant-form one, see, e.g., \cite{KP,Lev}).

 Now we use the observation which is the essence of the "minimal relativity
principle" \cite{BrKuo}:  if ${\tilde M}$ is the mass operator in ${\cal
H}_{int}'$ and the interaction operator $V$ in ${\cal H}_{int}'$ is defined as
${\tilde M}^2 = M_0^2+V$, then the equation ${\tilde M}^2\Psi_n = M_n^2\Psi_n$
for the eigenvalues $M_n$ and eigenfunctions $\Psi_n$ of the operator ${\tilde
M}$ has the same form as the nonrelativistic Schroedinger equation in momentum
representation:  \begin{equation} (\frac{{\vec k}^2}{m_0}+{\cal V})\Psi_n({\vec
k},\sigma_1,\sigma_2) = E_n\Psi_n({\vec k},\sigma_1,\sigma_2) \label{13}
\end{equation}

\noindent where

\begin{equation} {\cal V}=V/4m_0,\quad E_n=(M_n^2-4m_0^2)/4m_0 \label{14}
\end{equation}

\noindent The operator ${\cal V}$ should satisfy the same conditions as in the
nonrelativistic quantum mechanics:  the operator $({\vec k}^2/m_0)+{\cal V}$
should be selfadjoint and ${\cal V}$ should commute with ${\vec S}_{nr}$.

 We can formally introduce the operator ${\vec r}=\imath \partial /\partial
{\vec k}$ which is canonically conjugated with ${\vec k}$.  It is well known
that in the relativistic case there is no operator which has all the properties
of the position operator.  Therefore ${\vec r}$ has not all the properties of
the operator of the relative radius-vector between particles 1 and 2; it has
such properties only in the nonrelativistic or classical limit.  Nevertheless we
can choose the operator ${\cal V}$ as the operator of multiplication by a
function ${\cal V}(r)$, where $r=|{\vec r}|$.  In particular the function ${\cal
V}(r)$ can be chosen in such a way that the operator ${\tilde M}$ (and hence
$M$) has only a discrete spectrum, i.e., the property which in CQMs is
associated with confinement.

 We choose the function ${\cal V}(r)$ in the form ${\cal V}(r)=a^4r^2/m_0$,
where $a$ is some constant with the dimension $GeV$.  Then Eq.  (\ref{13}) is
the well-known equation for the harmonic oscillator.  We stress, once more, that
in Eq.  (\ref{13}) the mass operator is a fully relativistic one, though it
coincides with the Schroedinger operator.  The solutions of Eq.  (\ref{13}) are
given by the products $\Psi_{\vec {n},\sigma_{1}', \sigma_{2}'}$ $ = \psi_{\vec
{n}} ({\vec k}) \varphi_{\sigma_{1}',\sigma_{2}'}(\sigma_1,\sigma_2)$, where
$\varphi_{\sigma_{1}',\sigma_{2}'}(\sigma_1,\sigma_2)$ is the spin eigenfunction
of particles 1 and 2 with eigenvalues $\sigma_{1}',\sigma_{2}'$, respectively,
and \begin{eqnarray} &&\psi_{\vec n} ({\vec k}) =
\psi_{n_x}(k_x)\psi_{n_y}(k_y)\psi_{n_z}(k_z), \quad \vec {n} \equiv
(n_x,n_y,n_z), \nonumber\\ &&\psi_{n_i}(k_i) = (\frac{2\sqrt{\pi}}{a})^{1/2}
\Phi_{n_i}(\frac{k_i}{a}), \nonumber\\ &&\Phi_{n_i}(t) =
\frac{1}{(2^{n_i}n_i!)^{1/2}} exp(-\frac{t^2}{2}) H_{n_i}(t), \nonumber\\ &&i =
x,y,z.  \label{15} \end{eqnarray}

\noindent In Eq.  (\ref{15}) $n_i = 0,1,2...$, and $H_{n_i}$ is the Hermite
polynomial of the $n_i$-th order.  The normalization of $\psi_{n_i}$ is
$\int_{-\infty}^{\infty} |\psi_{n_i}(k_i)|^2~\frac{d{k_i}}{2\pi} = 1$.  As
follows from Eqs.  (\ref{13}) and (\ref{14}), the eigenvalues of the mass
operator are equal to \begin{equation} M_n=2[m_0^2+a^2(2n+3)]^{1/2} \label{16}
\end{equation}

\noindent with $n=n_x+n_y+n_z$.

 In order to calculate the DIS structure functions we have to know the wave
function of the two-body system in the space $\cal H$ when the internal wave
function $\chi_{\vec {n},\sigma_{1}',\sigma_{2}'}$ in the space ${\cal H}_{int}$
is related to $\Psi_{\vec {n},\sigma_{1}',\sigma_{2}'}$ according to Eq.
(\ref{12}) and the system as a whole is in the eigenstate of the operators
${\vec P}_{\bot}$ and $P^+$ with the eigenvalues ${\vec {P'}}_{ \bot}$ and
$P'^{+}$, respectively.  Let us first normalize as follows the one-particle
eigenstates with the four-momentum $p'$ and spin projection $\sigma'$:
\begin{sloppypar} \begin{equation} \langle
p",\sigma"|p',\sigma'\rangle=2(2\pi)^3p^{'+} \delta^{(2)}({\vec
{p}}^{\,"}_{\bot}- {\vec {p'}}_{\bot})
\delta(p^{"+}-p^{'+})\delta_{\sigma"\sigma'} \label{18} \end{equation}

\noindent ($\delta_{\sigma"\sigma'}$ is the usual Kronecker symbol).  Then the
state $|{\vec {P'}}_{\bot},P'^{+},\chi_{\vec {n},
\sigma_{1}^{'},\sigma_{2}^{'}}\rangle$ is described by the wave function
\end{sloppypar}

\begin{eqnarray} &&\langle {\vec P"}_{ \bot},P"^{+},{\vec k},\sigma_1,\sigma_2
|{\vec {P'}}_{ \bot},P'^{+},\chi_{\vec {n},\sigma_{1}^{'},\sigma_{2}^{'}}\rangle
= 2(2\pi)^3 \cdot \nonumber \\ &&P'^{+}\delta^{(2)}({\vec P"}_{ \bot}-{\vec
{P'}}_{ \bot}) \delta(P"^+-P'^{+}) \chi_{\vec {n},\sigma_{1}',\sigma_{2}'}
({\vec k},\sigma_1,\sigma_2) \label{19} \end{eqnarray}

From Eqs.  (\ref{12}) and (\ref{15}) it is clear that $||\chi_{\vec n,
\sigma_1,\sigma_2}||=1$, if the spin eigenfunctions are properly normalized.

\begin{sloppypar} \section{DIS structure functions in the parton model}
\label{S3} \end{sloppypar}

 Let us consider in some detail how the structure functions can be obtained
within the parton model in our light-cone framework.  The results are well known
(see, e.g., Ref.  \cite{Coester}), but the purpose of the following derivation
is to make a comparison with the one given in Sec.  4, where the final states
are described by the exact harmonic oscillator wave functions.  For simplicity
we will consider the case where the total spin of the initial system is $S=0$
(this is by no way a restriction:  indeed in the case where $S=1$ one can obtain
the same results as well).  Then the internal wave function of the initial state
will be \begin{eqnarray} &&\chi_{0}({\vec k},\sigma_1,\sigma_2) = \langle {\vec
k},\sigma_1,\sigma_2|\chi_{0}\rangle\nonumber\\ && = \omega({\vec k})^{1/2}
\psi_0({\vec k}) \langle \sigma_1,\sigma_2| U^{-1}|\varphi_{S=0}\rangle\
\label{19a} \end{eqnarray}

Let $J^{\mu}(x)$ be the electromagnetic current operator for the system under
consideration $(\mu=0,1,2,3)$, where $x$ is a point in Minkowski space.  It is
well known that if the initial state is $|{\vec {P'}}_{
\bot},P'^{+},\chi_0\rangle$ then the DIS hadronic tensor is given by

\begin{eqnarray} &&W^{\mu\nu} = \frac{1}{4\pi}\int\nolimits exp(\imath qx) \cdot
\nonumber\\ &&\cdot \langle {\vec {P'}}_{ \bot},P'^{+},\chi_{0}
|J^{\mu}(x)J^{\nu}(0) |{\vec {P'}}_{ \bot},P'^{+},\chi_{0}\rangle d^4x
\label{ht} \end{eqnarray}

\noindent The coordinate dependence of the current operator is fully defined by
translational invariance, according to which

\begin{equation} J^{\mu}(x)=exp(\imath Px)J^{\mu}(0)exp(-\imath Px) \label{tr}
\end{equation}

\noindent where $P$ is the total four-momentum operator.  Therefore, as follows
from Eqs.  (\ref{ht}) and (\ref{tr}), \begin{eqnarray}
&&W^{\mu\nu}=\frac{1}{4\pi} \sum_{X} (2\pi)^4\delta^{(4)}(P'+q-P_X) \cdot
\nonumber\\ &&\langle {\vec {P'}}_{ \bot},P^{'+},\chi_{0}| J^{\mu}(0) |X\rangle
\langle X |J^{\nu}(0)|{\vec {P'}}_{ \bot},P^{'+},\chi_{0}\rangle \label{20}
\end{eqnarray}

\noindent where a sum is taken over all possible final states $|X\rangle$ and
$P_X$ is the four-momentum of the state $|X\rangle$.  It is also well known
that, as a consequence of Poincar\'e invariance and current conservation, the
unpolarized hadronic tensor has the form \begin{eqnarray}
&&W^{\mu\nu}(P',q)=(\frac{q^{\mu}q^{\nu}}{q^2}-g^{\mu\nu}) F_1(x,Q)+\nonumber\\
&&\frac{1}{(P'q)}(P^{'\mu}-\frac{q^{\mu}(P'q)}{q^2})
(P^{'\nu}-\frac{q^{\nu}(P'q)}{q^2})F_2(x,Q), \label{21} \end{eqnarray}

\noindent where $g^{\mu\nu}$ is the Minkowski tensor.

In the parton model one assumes that in Bjorken limit the operator $J^{\mu}(0)$
can be taken in the impulse approximation, i.e., $J^{\mu}(0) =
\sum_{i=1}^{2}J_i^{\mu}(0) $ where the current operator $J_i^{\mu}(0)$ acts as
follows \begin{eqnarray} \langle p_i,\sigma_i| J_i^{\mu}(0)
|p_i',\sigma_i'\rangle &=& {\bar w}(p_i,\sigma_i)\gamma
^{\mu}w(p_i',\sigma_i')\nonumber\\ \label{22} \end{eqnarray}

\noindent where $\gamma^{\mu}$ are the Dirac $\gamma$ matrices and
$w(p_i,\sigma_i)$ is the light-cone Dirac spinor.

The following representation for the $\gamma$ matrices has been adopted

 \begin{equation} \gamma^0=\left\|\begin{array}{cc} 0 & 1\\ 1 & 0
\end{array}\right\|,\quad \gamma^5=\left\|\begin{array}{cc} 1 & 0\\ 0 & -1
\end{array}\right\|,\quad \gamma^i=\left\|\begin{array}{cc} 0 & -\tau_i\\ \tau_i
& 0 \end{array}\right\| \label{22'} \end{equation} where $i=1,2,3$ and $\tau_i$
are the Pauli matrices.  Then the light-cone Dirac spinor can be written as

\begin{equation} w(p,\sigma)=\sqrt{m_0} \left\|\begin{array}{c} \beta({\vec
g}_{\bot},g^+)\chi(\sigma)\\ \left [\beta({\vec g}_{\bot},g^+)^{-1} \right ]
^{\dagger}\chi(\sigma) \end{array}\right\| \label{23} \end{equation}

\noindent where $\chi(\sigma)$ is the ordinary spinor describing the state with
the spin projection on the $z$ axis equal to $\sigma$ and the matrix
$\beta({\vec g}_{\bot},g^+)$ (where $g=p/m_0$) has the components
\begin{equation} \beta_{11}=\beta_{22}^{-1}=2^{1/4}(g^+)^{1/2},\quad
\beta_{12}=0,\quad \beta_{21}=(g_x+\imath g_y)\beta_{22} \label{24}
\end{equation}

For simplicity we assume that particle 1 has unit electric charge and particle 2
is chargeless.  With this assumption, from Eqs.  (\ref{4}), (\ref{19}) and
(\ref{22}), we have \begin{eqnarray} &&\langle p_1",\sigma_1"| \langle
p_2",\sigma_2"| J^{\mu}(0) |{\vec {P'}}_{ \bot},P'^{+},\chi_{0} \rangle =
\nonumber\\ &=& \frac{1}{\xi'}\sum_{\sigma_1'} [{\bar
w}(p_1",\sigma_1")\gamma^{\mu}w(p_1',\sigma_1')] \chi_{0} ({\vec
k}',\sigma_1',\sigma_2") \label{26} \end{eqnarray}

\noindent where $\xi'$ and ${\vec k}'$ are constructed by means of Eqs.
(\ref{4},\ref{5}) from the vectors $p_1'$ and $P'$, where ${\vec {p '}}_{
1\bot}= {\vec {P'}}_{ \bot}-{\vec p^{\,"}}_{ 2\bot}$,
$p_1'^{+}=P'^{+}-p_2^{"+}$.

 As noted above, one of the major parton-model assumptions is that the FSI of
the struck quark with the target remnants can be neglected.  This implies that
in our case the final states $X$ in Eq.  (\ref{20}) are the states of two free
particles:  \begin{equation} |X\rangle =|p_1",\sigma_1"\rangle
|p_2",\sigma_2"\rangle \label{27} \end{equation}

\noindent and therefore Eq.  (\ref{20}) for the hadronic tensor can be written
as \begin{eqnarray} W^{\mu\nu}&=&\frac{1}{4\pi} \sum_{\sigma_1"\sigma_2"}
\int\nolimits (2\pi)^4 \delta^{(4)}(P'+q-P") \cdot \nonumber\\ &&\langle {\vec
{P'}}_{ \bot},P'^{+},\chi_{0}| J^{\mu}(0)|p_1^",\sigma_1",p_2^",\sigma_2"\rangle
\cdot\nonumber\\ &&\langle p_1^",\sigma_1",p_2^",\sigma_2"|J^{\nu}(0)| {\vec
{P'}}_{ \bot},P'^{+},\chi_{0} \rangle \cdot\nonumber\\ &&\prod_{i=1}^{2}
d\rho_i({\vec p}_{i\bot}^{"},p_i^{"+}) \label{28} \end{eqnarray}

\noindent where $P"=p_1^"+p_2^"$.  In turn, as follows from Eqs.  (\ref{6}),
(\ref{18}), (\ref{26}) and (\ref{27}), Eq.  (\ref{28}) can be written as
\begin{eqnarray}
&&W^{\mu\nu}=\frac{1}{4(P'^{+}+q^+)}\sum_{\sigma_1"\sigma_1'\sigma_2"}
\int\nolimits\delta(P'^{-}+q^--P"^{-})\cdot\nonumber\\ &&\chi_0({\vec
k}_{\bot}',\xi',\sigma_1",\sigma_2")^* [{\bar w}(p_1',\sigma_1")\gamma^{\mu}
({\not \!  p}_1"+m_0) \gamma^{\nu}w(p_1',\sigma_1')] \cdot\nonumber\\
&&\chi_0({\vec k}_{\bot}',\xi',\sigma_1',\sigma_2") \frac{d{\vec {k}}^{\,"}_{
\bot}d\xi"}{2(2\pi)^3(\xi')^2\xi"(1-\xi")} \label{29} \end{eqnarray}

\noindent where ${\not \!  p}=p^{\mu}\gamma_{\mu}$ and, as follows from Eqs.
(\ref{4}) and (\ref{5}), \begin{eqnarray} {\vec {p'}}_{ 1\bot}&=&{\vec
k}_{\bot}'+\xi'{\vec {P'}}_{ \bot}, \quad {\vec {p}}^{\,"}_{ 1\bot}={\vec
{p'}}_{ 1\bot}+{\vec q}_{\bot}= {\vec {k}}^{\,"}_{ \bot}+\xi"({\vec {P'}}_{
\bot}+{\vec q}_{\bot}), \nonumber\\ p_1'^{+}&=&\xi'P'^{+},\quad
p_1^{"+}=p_1'^{+}+q^+=\xi"(P'^{+}+q^+), \nonumber\\ {\vec {p}}^{\,"}_{
2\bot}&=&-{\vec {k}}^{\,"}_{ \bot}+(1-\xi")({\vec {P'}}_{ \bot}+ {\vec
q}_{\bot}),\nonumber\\ p_2^{"+}&=&(1-\xi')P'^{+}=(1-\xi")(P^{'+}+q^+) \label{30}
\end{eqnarray}

\noindent Therefore the relation between $({\vec k}_{\bot}',\xi')$ and $({\vec
{k}}^{\,"}_{ \bot},\xi")$ is \begin{equation} {\vec {k}}^{\,"}_{ \bot}={\vec
k}_{\bot}'+(\xi'-\xi"){\vec {P'}}_{ \bot}+ (1-\xi"){\vec q}_{\bot},\quad
\xi"(P^{'+}+q^+)=\xi'P^{'+}+q^+ \label{31} \end{equation}

\begin{sloppypar} In the parton model the initial hadron is considered in the
 infinite momentum frame (IMF), i.e., in a reference frame where $P'^{z}$ is
 positive and very large.  It is also assumed that the transverse momenta of the
 constituents in the initial and final states are restricted by some value (say
 $300 MeV/c$).  In view of this, a suitable choice of the reference frame is
 such that ${\vec {P'}}_{ \bot}= {\vec q}_{\bot}=0$.  Then, from Eqs.
 (\ref{30}) and (\ref{31}), one has \end{sloppypar}

 \begin{equation} {\vec {p}}^{\,"}_{ 1\bot}={\vec {p'}}_{ 1\bot}={\vec
{k}}^{\,"}_{ \bot}={\vec {k'}}_{ \bot}, \quad {\vec {p}}^{\,"}_{ 2\bot}=-{\vec
{k'}}_{ \bot} \label{32} \end{equation}

\begin{sloppypar} The above conditions do not define the reference frame
uniquely, since one can still boost this frame along the $z$ axis and choose the
Breit frame, where $\vec{P"} + \vec{P'} = 0$, in order to study DIS (let us
recall that such boosts along the $z$ axis are kinematical in the front form).
It is important to point out that, as shown in Ref.  \cite{LPS}, in the Breit
frame the one-body current operator, Eq.  (\ref{26}), is fully compatible with
the Poincar\'e trasformation properties.

 In the chosen frame, as follows from the definition of the Bjorken variable
$x$, one has in the Bjorken limit \begin{eqnarray} &&q^0=2P_z'(1-x),\quad
P^{'+}=\sqrt{2}P_z',\quad q^+=-\sqrt{2}P_z'x, \nonumber\\
&&P^{"+}=\sqrt{2}P_z'(1-x) \label{Breit} \end{eqnarray}

\noindent and the relation between $\xi'$ and $\xi"$ (see Eq.  (\ref{31}))
becomes \begin{equation} \xi'=x+(1-x)\xi" \label{33} \end{equation}
\end{sloppypar}

 Furthermore, by a direct calculation using Eq.  (\ref{Breit}), one obtains that
the relevant elements of the hadronic tensor (\ref{21}) are given in the Bjorken
limit by the following expressions \begin{eqnarray}
&&W^{jl}=\delta^{jl}F_1(x,Q),\quad (j,l=1,2) \nonumber\\ &&
W^{++}=\frac{1}{4(2-x)}[F_2(x,Q)-2xF_1(x,Q)],\nonumber\\ &&W^{+-}=W^{-+}=
\frac{1}{4x}[F_2(x,Q)-2xF_1(x,Q)],\nonumber\\
&&W^{--}=\frac{2-x}{4x^2}[F_2(x,Q)-2xF_1(x,Q)] \label{33'} \end{eqnarray}

\begin{sloppypar} Finally, in the Bjorken limit the argument of the
 delta-function in Eq.  (\ref{29}) becomes proportional to \end{sloppypar}

$$\frac{Q^2(1-x)}{x}-\frac{(m_0^2+{\vec k}_{\bot}^{"2})}{\xi" (1-\xi")}$$

\noindent Therefore we have two solutions for $\xi"$, the first of which is
close to 0 and the second is close to 1.  A direct calculation based on Eqs.
(\ref{23}), (\ref{24}), (\ref{29}), (\ref{32}) and (\ref{33}) shows that in the
Bjorken limit the contribution of the second solution is negligible.  Indeed,
when $\xi"$ is close to 1, then $\xi'$ is close to one too (see Eq.  (\ref{33}))
and $k_z'$ goes to infinity (see Eq.  (\ref{5})), making vanishing $\chi_0$ in
Eq.  (\ref{29}).  In what follows we describe only the contribution of the first
solution ($\xi"$ close to 0).

 The only large component of the momentum of particle 1 in the initial state is
$p_1'^{+}$ and the only large component of the momentum of this particle in the
final state is $p_1^{"-}$.  Then, as easily seen from Eq.  (\ref{29}), the spin
structure of the tensor $W^{\mu\nu}$ in the Bjorken limit is \begin{equation}
W^{\mu\nu}\sim Tr(\gamma^-\gamma^{\mu}\gamma^+\gamma^{\nu}) \label{33"}
\end{equation}

\noindent where $Tr$ stands for trace.  This implies that all the longitudinal
components of $W^{\mu\nu}$ are equal to zero, i.e., \begin{equation} W^{+\nu} =
W^{-\nu} = W^{\mu+ }= W^{\mu-} = 0, \label{35'} \end{equation}

\noindent since $(\gamma^+)^2 = (\gamma^-)^2 = 0$.  The only non-vanishing
components are the transverse ones, viz.  \begin{eqnarray}
W^{jl}&=&\sum_{\sigma_1"\sigma_1'\sigma_2"}\int\nolimits \chi_0({\vec
k}_{\bot}',x,\sigma_1",\sigma_2")^*
[\delta_{jl}\delta_{\sigma_1"\sigma_1'}+2\imath \epsilon_{jl}
(s_1^z)_{\sigma_1"\sigma_1'}]\cdot\nonumber\\ &&\chi_0({\vec
k}_{\bot}',x,\sigma_1',\sigma_2") \frac{d{\vec k}_{\bot}'}{4(2\pi)^3x(1-x)}
\label{34} \end{eqnarray}

\noindent In obtaining Eq.  (\ref{34}) we have used the fact that, if $\xi"$ is
close to 0, then in the Bjorken limit $\xi'=x$ (see Eq.  (\ref{33})).  As
follows from Eq.  (\ref{4}), the quantity $\xi'$ is the momentum fraction of
particle 1 in the initial state.  Therefore we obtain the well-known result that
the Bjorken variable $x$ has the meaning of the momentum fraction of the struck
quark in the IMF.

 The structure function $F_1(x,Q)$ is given by the symmetrical part of $W^{jl}$
and the structure function $F_2(x,Q)$ by the longitudinal components.  Let us
introduce the notation \begin{eqnarray}
\rho(x)&=&\sum_{\sigma_1,\sigma_2}\int\nolimits |\chi_0({\vec
k}_{\bot},x,\sigma_1,\sigma_2)|^2 \frac{d{\vec k}_{\bot}}{2(2\pi)^3x(1-x)} =
\nonumber\\ &=&\int\nolimits |\chi_0({\vec k}_{\bot},x)|^2 \frac{d{\vec
k}_{\bot}}{2(2\pi)^3x(1-x)} \label{35} \end{eqnarray}

\noindent where (see Eqs.  (\ref{12}) and (\ref{15})) $\chi_0({\vec
k}_{\bot},\xi) = \omega({\vec k})^{1/2} \psi_0({\vec k})$.  Then, as follows
from Eqs.  (\ref{4}), (\ref{6}) and (\ref{10}), $\rho(x)dx$ is the probability
to have a momentum fraction of particle 1, in the initial state, falling in the
interval $(x,x+dx)$.  By using Eqs.  (\ref{33'}), (\ref{35'}), (\ref{34}) and
(\ref{35}), one obtains \begin{equation} F_1(x) = \frac{1}{2}\rho(x) \label{36}
\end{equation}

\noindent and the Callan-Gross relation \cite{CallGr} \begin{equation} F_2(x) =
2xF_1(x) \label{36'} \end{equation}

\noindent Therefore in the Bjorken limit the structure functions $F_1$ and $F_2$
do not depend on $Q$, namely one has Bjorken Scaling.

\begin{sloppypar} \section{DIS structure functions with the relativistic
 harmonic oscillator wave functions} \label{S4} \end{sloppypar}

 Let us consider the exact hadronic tensor for two particles, interacting via
the relativistic harmonic oscillator potential introduced in Sec.  2, and let
the initial state be the $S = 0$ state as in Sec.  3.  Therefore in Eq.
(\ref{20}) we use, as final states $|X\rangle$, the {\em exact eigenfunctions}
defined by Eqs.  (\ref{12}), (\ref{15}), (\ref{19}).  Then the exact expression
for the hadronic tensor of the system under consideration is

\begin{eqnarray} &&W^{\mu\nu}=\frac{1}{4\pi} \sum_{\sigma_1"\sigma_2"} \sum_{n_x
n_y n_z}\int\nolimits (2\pi)^4 \delta^{(4)}(P'+q-P"_n) \cdot \nonumber\\ &&
\langle {\vec {P'}}_{ \bot},P'^{+},\chi_{0}| J^{\mu}(0)|{\vec {P}}^{\,"}_{
\bot},P"^{+},\chi_{\vec {n},\sigma_{1}",\sigma_{2}"} \rangle \cdot \nonumber\\
&& \langle {\vec {P}}^{\,"}_{ \bot},P"^{+},\chi_{\vec
{n},\sigma_{1}",\sigma_{2}"} |J^{\nu}(0)| {\vec {P'}}_{ \bot},P'^{+},\chi_{0}
\rangle \cdot \nonumber\\ && d\rho({\vec {P}}^{\,"}_{ \bot},P"^{+}) \label{37}
\end{eqnarray}

\noindent where the four-vectors $P"_n = P' + q$ have the components
$P"_n^{+}=P"^{+}$, ${\vec P"}_{ n\bot}={\vec {P}}^{\,"}_{ \bot}$, $P"_n^{-}=
(M_n^2+{\vec {P}}^{\," 2}_{ \bot})/2P"^{+}$, \begin{eqnarray} &&|\chi_{\vec
{n},\sigma_1",\sigma_2"}({\vec k}) \rangle= \omega({\vec k})^{1/2} \psi_{\vec
n}({\vec k}) U^{-1} |\varphi_{\sigma_1",\sigma_2"}\rangle , \label{38}
\end{eqnarray}

\noindent and $|\varphi_{\sigma_1,\sigma_2}\rangle$ is the normalized spin
eigenstate of particles 1 and 2 with the eigenvalues $\sigma_1,\sigma_2$,
respectively:  \begin{equation} ||\varphi_{\sigma_1,\sigma_2}|| = 1.  \label{39}
\end{equation}

\begin{sloppypar} We will calculate the tensor (\ref{37}) in the same reference
frame as in the preceding section, i.e., in the Breit frame (${\vec P}" + {\vec
P}' = 0$) with ${\vec {P'}}_{ \bot} = {\vec q}_{\bot} = 0$.  We assume
$J^{\mu}(0) = J_1^{\mu}(0)$ for $\mu~=~+,1,2$ and use current conservation to
define $J_1^-(0)$.  As it is shown in \cite{LPS}, this current operator is an
allowed choice, which is fully compatible in our reference frame with Poincar\'e
transformation properties, and trivially fulfills current conservation.  Then,
from Eqs.  (\ref{1}), (\ref{6}), (\ref{19}), (\ref{21}) and (\ref{26}),

\begin{eqnarray} &&F_1(x,Q)=\frac{1}{2}\sum_{\sigma_1"\sigma_2"} \sum_{n_x n_y
n_z}\delta(m^2+\frac{Q^2(1-x)}{x}-M_n^2) \cdot \nonumber\\
&&|\sum_{\sigma_1\sigma_1'\sigma_2'} \int\nolimits \chi^+_{\vec
{n},\sigma_1",\sigma_2"}({\vec {k}}^{\,"}_{ \bot},\xi",\sigma_1,\sigma_2')
[{\bar w}(p_1^",\sigma_1)\gamma^1w(p_1',\sigma_1')] \cdot \nonumber\\ &&\chi_{0}
({\vec k}_{\bot}',\xi',\sigma_1',\sigma_2') \frac{d{\vec {k}}^{\,"}_{
\bot}d\xi"}{2(2\pi)^3\xi'\xi" (1-\xi")}|^2 \label{40} \end{eqnarray}

\noindent where $m=2[m_0^2+3a^2]^{1/2}$ is the ground state eigenvalue of the
mass operator and the relations between the quantities $p_1', p_1^", {\vec
k}_{\bot}', \xi', {\vec {k}}^{\,"}_{ \bot}, \xi"$ are given by Eqs.  (\ref{30})
and (\ref{31}).  In the reference frame under consideration, ${\vec {k}}^{\,"}_{
\bot}={\vec k}_{\bot}'$ as in the parton model, and the relation between $\xi"$
and $\xi'$ is still given by Eq.  (\ref{33}).  \end{sloppypar}

The spin sums in Eq.  (\ref{40}) can be easily done taking into account the
following expressions for the Clebsh-Gordan coefficient in the initial state,
the Melosh rotation and the matrix element of the parton current, viz.
\begin{eqnarray} && \langle \sigma_1 \sigma_2|00 \rangle =i\langle \sigma_1
|{\sigma_y \over \sqrt{2}}|\sigma_2 \rangle \nonumber \\ && v(\vec{k},\vec{s})
={ m_0+ \omega(\vec{k})+k_z-i 2\vec{s} \cdot (\hat{e}_z \times \vec{k}_{\perp})
\over \sqrt{2(\omega(\vec{k}) +m_0) (\omega(\vec{k}) +k_z)}} \nonumber \\
&&\bar{w}(p_1^",\sigma_1) \gamma_x w(p_1', \sigma_1')= \nonumber \\ &&{1 \over
\sqrt{p_1"^+p_1'^+}} \langle \sigma_1 | \left [ q^+ (i \sigma_y m+k_x"+i
k_y"\sigma_z) + 2p_1'^+ k_x" \right ] |\sigma_1' \rangle \label{40a}
\end{eqnarray} Then Eq.  (\ref{40}) becomes \begin{eqnarray}
&&F_1(x,Q)=\frac{1}{2} \sum_{n_xn_yn_z} \delta(m^2+\frac{Q^2(1-x)}{x}-M_n^2)
\cdot \nonumber\\ &&\left [| \int\nolimits \chi_{\vec n}({\vec k}_{\perp},\xi)
{1 \over \sqrt{p_1"^+p_1'^+}}q^+ \left [m_0+{k_y^2 \over \omega({\vec k})+m_0}
\right ] ~ \chi_{\vec{0}} ({\vec k}_{\perp},\xi') \frac{d{\vec {k}}_{ \bot}
d\xi}{2(2\pi)^3\xi'\xi (1-\xi)}|^2 + \right.  \nonumber \\ &&| \int\nolimits
\chi_{\vec n}({\vec k}_{\perp},\xi) {1 \over \sqrt{p_1"^+p_1'^+}}{q^+k_yk_z\over
[\omega({\vec k})+m_0]} ~\chi_{\vec{0}} ({\vec k}_{\perp},\xi') \frac{d{\vec
{k}}_{ \bot}d\xi}{2(2\pi)^3\xi'\xi (1-\xi)}|^2 +\nonumber \\ && | \int\nolimits
\chi_{\vec n}({\vec k}_{\perp},\xi) {1 \over
\sqrt{p_1"^+p_1'^+}}(q^++2p_1^{'+})k_x ~\chi_{\vec{0}} ({\vec
k}_{\perp},\xi')\frac{d{\vec {k}}_{ \bot}d\xi}{2(2\pi)^3\xi'\xi (1-\xi)}|^2
+\nonumber \\ && \left .  | \int\nolimits \chi_{\vec n}({\vec k}_{\perp},\xi) {1
\over \sqrt{p_1"^+p_1'^+}}\frac{q^+k_yk_x}{[\omega({\vec k})+m_0]}
~\chi_{\vec{0}} ({\vec k}_{\perp},\xi') \frac{d{\vec {k}}_{
\bot}d\xi}{2(2\pi)^3\xi'\xi (1-\xi)}|^2 \right ] \label{40b} \end{eqnarray}
\noindent where in order to simplify the notation, the variables ${\vec
{k}}^{\,"}_{ \bot}$ and $\xi"$ of Eq.  (\ref{40}) have been replaced by ${\vec
k}_{\bot}$ and $\xi$; moreover (see Eqs.  (\ref{33}) and (\ref{38}))

\begin{equation} \chi_{\vec n}({\vec k})= \omega({\vec k})^{1/2}\psi_{\vec
n}({\vec k}),\quad \chi_0({\vec k}')=\omega({\vec k}')^{1/2}\psi_0({\vec
k}'),\quad \xi'=x+(1-x)\xi, \label{42} \end{equation}

\noindent and the relation between ${\vec k}$ (${\vec k}'$) and $({\vec
k}_{\bot},\xi)$ ($({\vec k}'_{\bot},\xi')$) is given by Eq.  (\ref{5}).

\medskip

\subsection {A pedagogical example}

In order to begin with a simple, pedagogical model that allows us to introduce
the mathematical tools for the general case (to be considered below), we assume
for the moment $a \ll m_0$.  This assumption implies that the relevant momenta
${\vec k}'$ in the wave function of the ground state satisfy the condition
$|{\vec k}'|\ll m_0$ (see Eq.  (\ref{15})), but does not destroy the
relativistic nature of the highly excited final states.  Then Eq.  (\ref{40b})
greatly simplifies, putting in evidence the relevant integral over $\xi$.  With
the help of Eq.  (\ref{Breit}), $F_1(x,Q)$ becomes \begin{eqnarray}
&&F_1(x,Q)=\frac{m_0^2x^2}{2(1-x)}\sum_{n_xn_yn_z}
\delta(m^2+\frac{Q^2(1-x)}{x}-M_n^2) \cdot\nonumber\\ &&|\int\nolimits
\chi_{\vec n}({\vec k}_{\bot},\xi) \chi_0({\vec k}_{\bot},\xi')\frac{d{\vec
k}_{\bot}d\xi} {2(2\pi)^3(\xi'\xi)^{3/2}(1-\xi)}|^2 \label{41} \end{eqnarray}

 So far we have not used the explicit expressions for the harmonic oscillator
wave functions and therefore, if the FSI is neglected in Eq.  (\ref{41}), the
result for $F_1(x,Q)$ should be the same as in Eq.  (\ref{36}).  Disregarding
the FSI in Eq.  (\ref{41}) implies the following replacements

\begin{eqnarray} \sum_{n_xn_yn_z}& \rightarrow &\int\nolimits \frac{d{\vec
{k}}^{\,"}_{ \bot}d\xi"}{2(2\pi)^3\xi"(1-\xi")}, \quad M_n^2 \rightarrow
\frac{m_0^2+{\vec k}_{\bot}^{"2}}{\xi"(1-\xi")},\nonumber\\ \chi_{\vec n}({\vec
k}_{\bot},\xi) &\rightarrow &2(2\pi)^3\xi"(1-\xi") \delta^{(2)}({\vec
k}_{\bot}-{\vec {k}}^{\,"}_{ \bot})\delta(\xi-\xi") \label{43} \end{eqnarray}

\noindent Then it is easy to see that we again arrive at Eq.  (\ref{36}) for
$F_1(x,Q)$.

 Let us go back to Eq.  (\ref{41}), where the {\em exact final state wave
functions} are used.  Since only the values $|{\vec k}_{\bot}| \ll m_0$ are
important in Eq.  (\ref{41}), we can neglect ${\vec k}_{\bot}$ in $\omega({\vec
k})$.  Then, as follows from Eqs.  (\ref{5}), (\ref{15}) and (\ref{42}), the
integration over ${\vec k}_{\bot}$ in Eq.  (\ref{41}) is trivial and as a result
\begin{eqnarray} &&F_1(x,Q)=\frac{m_0^2x^2}{2(1-x)}\sum_{n}
\delta(m^2+\frac{Q^2(1-x)}{x}-M_n^2)\cdot\nonumber\\ &&|\int_{0}^{1}
(m_0^2+k_z^2)^{1/4} \psi_n(k_z)\chi_0(\xi')\frac{d\xi}
{4\pi(\xi'\xi)^{3/2}(1-\xi)}|^2 \label{44} \end{eqnarray}

\noindent where $n$ replaces $n_z$, $\chi_0(\xi') = m_0^{1/2}\psi_0(k_z')$, the
relation between $\xi$ and $\xi'$ is given by Eq.  (\ref{42}) and one-to-one
relations $\xi \leftrightarrow k_z$ and $\xi' \leftrightarrow k_z'$ can be
obtained from Eq.  (\ref{5}).  In particular if the dependence of $\omega({\vec
k})$ on ${\vec k}_{\bot}$ is neglected, then the relation between $\xi$ and
$k_z$ is given by

\begin{eqnarray} &&\xi -\frac{1}{2}=\frac{k_z}{2(m_0^2+k_z^2)^{1/2}},\quad
k_z=\frac{m_0(\xi-1/2)}{[\xi(1-\xi)]^{1/2}},\nonumber\\
&&\frac{d\xi}{2\xi(1-\xi)}=\frac{dk_z}{(m_0^2+k_z^2)^{1/2}} \label{45}
\end{eqnarray}

\noindent and the relation between $\xi'$ and $k_z'$ is the same.

As a consequence of Eq.  (\ref{45}), Eq.  (\ref{44}) can be written in the form
\begin{eqnarray} F_1(x,Q)&=&\frac{m_0^2x^2}{8\pi^2(1-x)}\sum_{n}
\delta(m^2+\frac{Q^2(1-x)}{x}-M_n^2)\cdot\nonumber\\
&&|\int_{-\infty}^{\infty}\frac{\psi_n(k_z)\chi_0(\xi')dk_z}
{\xi'(\xi'\xi)^{1/2}(m_0^2+k_z^2)^{1/4}}|^2 \label{46} \end{eqnarray}

 This is the last stage where we still can return back to the parton model.
Indeed, as follows from Eqs.  (\ref{43}) and (\ref{44}), neglecting FSI in Eq.
(\ref{46}) implies the following replacements

\begin{eqnarray} &&\sum_{n} \rightarrow \int\nolimits
\frac{dk_z^"}{2\pi(m_0^2+k_z^{"2})^{1/2}}, \quad M_n^2 \rightarrow
4(m_0^2+k_z^{"2}), \nonumber\\ &&\psi_n(k_z) \rightarrow
2\pi(m_0^2+k_z^2)^{1/4}\delta(k_z-k_z^") \label{47} \end{eqnarray}

\noindent With these replacements it is easy to see that, in the Bjorken limit,
Eq.  (\ref{46}) leads to Eq.  (\ref{36}) for $F_1(x,Q)$, if $\chi_{\vec 0}({\vec
k})$ is given by Eq.  (\ref{42}) and $a\ll m_0$.

  In the last part of this section we will consider Eq.  (\ref{46}) with the
function $\psi_n(k_z)$ given by the {\em exact eigenfunction}, i.e.  by Eq.
(\ref{15}).

 As follows from Eqs.  (\ref{16}) and (\ref{46}), $n=Q^2(1-x)/8a^2x$ and
therefore, if $Q$ is large, only large values of $n$ are important in Eq.
(\ref{46}).  Taking into account Eq.  (\ref{16}) we can write Eq.  (\ref{46}) in
the form

\begin{equation} F_1(x,Q)=\frac{m_0^2x^4}{8\pi^2(1-x)Q^2}\sum_{n}
\delta(x-\frac{Q^2}{Q^2+8a^2n}) f(n,x)^2 \label{48} \end{equation}

\noindent where

\begin{equation} f(n,x)=(-1)^n\int_{-\infty}^{\infty}
\frac{\psi_n(k_z)\chi_0(\xi')dk_z}{\xi'(\xi'\xi)^{1/2} (m_0^2+k_z^2)^{1/4}}
\label{49} \end{equation}

\noindent We will show in Appendix A that $f(n,x)$ is a smooth function of $x$
and has a finite limit for $n \rightarrow \infty $.

Now the following question arises.  While in the parton model $F_1(x)$ and
$F_2(x)$ are continuous functions of $x$, it is clear from Eq.  (\ref{48}) that
at fixed $Q$ the function $F_1(x,Q)$ is a linear combination of delta-functions,
which are not equal to zero only for discrete values of $x$.  As noted in Sec.
\ref{S1}, the first impression is that a correspondence between discrete and
continuous cases cannot exist.  However, Eq.  (\ref{48}) is meaningful only in
the realm of the distributions.  Within such a framework, the correspondence
between the discrete and continuous cases could be shown in the Bjorken limit,
since the discrete values of $x$, where $F_1(x,Q)$ is not zero, become closer
and closer as $Q$ increases.

 In order to simplify the mathematical discussion, let us note that experiments
allow one to determine not the very function $F_1(x,Q)$, but its average values
over some bins in $x$ and $Q$.  Therefore, let us consider the integral

\begin{equation} {\bar F}_1({\bar x},Q)=\frac{1}{x_2-x_1}\int_{x_1}^{x_2}
F_1(x,Q)dx , \label{50} \end{equation}

\noindent where ${\bar x}$ belongs to the small interval $[x_1,x_2]$, such that
$x_2-x_1\ll {\bar x}$.  It is clear that for large Q there exist many values of
$n$ such that $Q^2/(Q^2+8a^2n) \in [x_1,x_2]$, even for a small, but finite
value of $x_2-x_1$.  Therefore the integral (\ref{50}) is a smooth function of
$Q$.  Let us define $n_1$ and $n_2$ as follows

\begin{equation} n_1={Q^2(1-x_1) \over 8a^2x_1}, \quad n_2={Q^2(1-x_2) \over
8a^2x_2}.  \label{51} \end{equation}

\noindent As a consequence $n_1-n_2= (x_2-x_1)n_2/[x_1(1-x_2)]\ll n_2$.  From
Eqs.  (\ref{48}) and (\ref{50}) one has

\begin{equation} {\bar F}_1({\bar x},Q)=\frac{m_0^2{\bar x}^4} {8\pi^2(1-{\bar
x})(x_2-x_1)Q^2}\sum_{n=n_2}^{n_1} f(n,{\bar x})^2 \label{52} \end{equation}

 The scaling property of the function ${\bar F}_1$ holds if $f(n,x)\rightarrow
g(x)$ when $n\rightarrow \infty$.  As a matter of fact, in this case the
quantity $\sum_{n=n_2}^{n_1} f(n,{\bar x})^2$ for large values of $n_1$ and
$n_2$ becomes $(n_1-n_2)g({\bar x})^2$ and therefore from Eq.  (\ref{51}) one
has

\begin{equation} {\bar F}_1({\bar x},Q)=\frac{m_0^2{\bar x}^2g({\bar x})^2}
{64a^2\pi^2(1-{\bar x})}.  \label{54} \end{equation}

\noindent Moreover, as follows from Eqs.  (\ref{35}), (\ref{36}) and (\ref{45}),
the result (\ref{54}) for the structure function will be equal to the parton
model one if \begin{equation} \lim_{n\rightarrow \infty} f(n,x) = g(x) =
\frac{(8\pi)^{1/2}a\chi_0(x)}{m_0x^{3/2}} \label{55} \end{equation}

\noindent where $\chi_0(x)=m_0^{1/2}\psi_0(k_z(x))$, with $k_z(x) = m_0 (x-1/2)
[x(1-x)]^{-1/2}$.  Indeed in our case, where the initial state is a harmonic
oscillator ground state with $a\ll m_0$, the structure function $F_1(x)$ in Eq.
(\ref{36}) becomes

\begin{equation} F_1(x) = \frac{\chi_0(x)^2}{8\pi x(1-x)}.  \label{55'}
\end{equation}

 In Appendix A it is shown how Eq.  (\ref{55}) can be proved.  In conclusion our
result for $F_1(x,Q)$ (Eq.  (\ref{48})) is indeed compatible with the parton
model, once an average over bins of $x$ is performed.

 As follows from the analysis of Appendix A (see in particular the discussion
about $f_2(n,x)$), the main contribution to the hadronic tensor is given by the
region where $k_z$ is negative and $|k_z|$ is very large (recall that
$t=k_z/a$).  In this region $\xi$ is small and $\rightarrow 0$ in the Bjorken
limit.  In turn, the quantity $\xi'$ is close to $x$ (see Eq.  (\ref{33})).  We
see that the usual interpretation of the Bjorken variable $x$ as the momentum
fraction of the struck quark is still valid if the FSI is not neglected.

 We recall that, with our choice of the reference frame, $|{\vec {p'}}_{
1\bot}|$ and $|{\vec {p}}^{\,"}_{ 1\bot}|$ can be neglected with respect to
$m_0$ (see Eq.  (\ref{30})) in the evaluation of the matrix elements of
$\gamma^\mu$ between light-cone Dirac spinors.  Therefore a direct calculation
using Eqs.  (\ref{23}) and (\ref{24}) shows that the matrix elements ${\bar
w}(p_1^",\sigma_1")\gamma^{+}w(p_1',\sigma_1')$, up to constant quantities, are
proportional to $\delta_{\sigma_1"\sigma_1'}(p^{"+} p^{'+})^{1/2}$.  Then, as
shown in Appendix C, $W^{++}$ is vanishing in the Bjorken limit, and from Eq.
(\ref{33'}) one obtains again that the Callan-Gross relation (\ref{36'}) holds.

 Finally, using once more Eq.  (\ref{33'}), one obtains that all the
longitudinal components of the hadronic tensor become negligible in the Bjorken
limit, with the exact wave functions for the final states as well as in the
parton model.

\medskip

\subsection {The general case}

 In the general case, i.e.  for any value of the harmonic oscillator strenght
$a$, the parton model can be recovered once again.  The starting point is Eq.
(\ref{40b}) that is exact.  After averaging over the $x$-bins, as in Eq.
(\ref{50}), one has \begin{eqnarray} {\bar F}_1({\bar x},Q)=\frac{{\bar x}^4}
{2(1-{\bar x})(x_2-x_1)Q^2}\sum_{n=n_2}^{n_1} ~\sum_{n_x+n_y+n_z=n}{\cal
F}(\vec{n},{\bar x}) \label{avF} \end{eqnarray}

\noindent where in the sum all the values of $n_i$ constrained by
$n=n_x+n_y+n_z$ are allowed, and $n_{1(2)}$ is defined in Eq.  (\ref{51}).  By
using in Eq.  (\ref{40b}) the expressions of $p_1'^+$ and $p_1"^+$ obtained from
Eqs.  (\ref{30}) and (\ref{Breit}), ${\cal F}(\vec{n},{\bar x})$ is given by
\begin{eqnarray} &&{\cal F}(\vec{n},{\bar x})=\left [| \int\nolimits \psi_{\vec
n}({\vec k}) {1 \over \sqrt{\xi\xi'}} \left [m_0+{k_y^2 \over \omega({\vec
k})+m_0} \right ] ~ \chi_{\vec{0}} ({\vec k}') {d{\vec {k}} \over (2\pi)^3 \xi'
\omega({\vec k})^{1/2}}|^2 + \right.  \nonumber \\ && | \int\nolimits \psi_{\vec
n}({\vec k}) {1 \over \sqrt{\xi\xi'}}{k_y k_z \over [\omega({\vec k})+m_0]}
~\chi_{\vec{0}} ({\vec k}') {d{\vec {k}} \over (2\pi)^3 \xi' \omega({\vec
k})^{1/2}}|^2 + \nonumber \\ && | \int\nolimits \psi_{\vec n}({\vec k}) {1 \over
\sqrt{\xi\xi'}}k_x(1 - 2 {\xi' \over \bar{x}}) ~\chi_{\vec{0}} ({\vec k}')
{d{\vec {k}} \over (2\pi)^3 \xi' \omega({\vec k})^{1/2}}|^2 +\nonumber \\ &&
\left .  | \int\nolimits \psi_{\vec n}({\vec k}) {1 \over \sqrt{\xi\xi'}}{k_y
k_x \over [\omega({\vec k})+m_0]} ~\chi_{\vec{0}} ({\vec k}') {d{\vec {k}} \over
(2\pi)^3 \xi' \omega({\vec k})^{1/2}}|^2 \right ] \label{f1} \end{eqnarray}

It is important to note that using the completeness over $n_x$ and $n_y$ one has
\be &&\sum_{n_x=0}^{\infty} \sum_{n_y=0}^{\infty}{\cal F}(\vec{n},\bar{x})= \int
{ d{\vec k}_{\perp} \over (2\pi)^2} |\psi_{0}(k_x)\psi_{0}(k_y)|^2 \cdot
\nonumber \\ &&\left [| \int\nolimits \psi_{n_z}(k_z) {1 \over
\sqrt{\xi\xi'}}\left [m_0+{k_y^2 \over [\omega({\vec k})+m_0]} \right ] ~
\chi_{0} ( k'_z) {{d k}_z \over 2\pi \xi' \omega({\vec k})^{1/2}}|^2 + \right.
\nonumber \\ &&| \int\nolimits \psi_{n_z}(k_z){1 \over \sqrt{\xi\xi'}}{k_y k_z
\over [\omega({\vec k})+m_0]} ~\chi_{0} ( k'_z) {{d k}_z \over 2\pi \xi'
\omega({\vec k})^{1/2}}|^2 +\nonumber \\ && | \int\nolimits \psi_{n_z}(k_z){1
\over \sqrt{\xi\xi'}}k_x(1 - 2 {\xi' \over \bar{x}}) ~\chi_{0} ( k'_z) {{d k}_z
\over 2\pi \xi' \omega({\vec k})^{1/2}}|^2 +\nonumber \\ && \left .  |
\int\nolimits \psi_{n_z}(k_z){1 \over \sqrt{\xi\xi'}}{k_y k_x \over
[\omega({\vec k})+m_0]} ~\chi_{0} ( k'_z) {{d k}_z \over 2\pi \xi' \omega({\vec
k})^{1/2}}|^2\right ], \label{f2} \ee

\noindent with $ \chi_{0} ( k'_z)=\omega(k'_z,{\vec k}_{\perp})^{1/2} \psi_{0} (
k'_z)$.  Then applying similar arguments as in Appendix A, where $m_0$ has to be
replaced by $\sqrt{m_0^2 +k^2_{\perp}}$, one obtains, as in the pedagogical
example, that only the negative and large values of $k_z$ contribute to the
integrals in Eq.  (\ref{f2}) in the Bjorken limit, and therefore $\xi'
\rightarrow x$.  Finally one can find \be \lim_{n_z \rightarrow \infty}
\sum_{n_x=0}^{\infty} \sum_{n_y=0}^{\infty}{\cal F}(\vec{n},\bar{x}) ={8 a ^2
\over \bar{x}^3} \int { d{\vec k}_{\perp} \over 2(2\pi)^3}|\chi_{0} ( {\vec
k}_{\perp},\bar{x})|^2 ={8 a ^2 \rho(\bar{x})(1-\bar{x})\over \bar{x}^2}
\label{f3} \ee

\noindent We explicitly note that in Eq.  (\ref{f3}) the order of $ \lim_ {n_z
\rightarrow \infty}$ and of the integration over $k_{\perp}$ has been exchanged,
thanks to the presence in Eq.  (\ref{f2}) of the gaussians
$\psi_{0}(k_x)\psi_{0}(k_y)$ and to the power-law behaviour in $\vec{k}_{\perp}$
of the remaining terms, for $k_{x(y)} \rightarrow \pm \infty$.  Equation
(\ref{f3}) implies that the positive function ${\cal F}(\vec{n},\bar{x})$ is
bounded by a quantity which does not depend upon $\vec{n}$.

Let us come back to Eq.  (\ref{avF}), that can be rewritten as

\begin{eqnarray} &&{\bar F}_1({\bar x},Q)=\frac{{\bar x}^2} {16 a^2 (1-{\bar
x})(n_1-n_2)}\sum_{n_x=0}^{n_1} \sum_{n_y=0}^{n_1-n_x} \cdot \nonumber \\
&&\sum_{n_z= MAX(0,n_2-n_x-n_y)}^{n_1-n_x-n_y}{\cal F}(\vec{n},{\bar x})
\label{avF0} \end{eqnarray} First of all, let us demonstrate that
\begin{eqnarray} \lim_{n_x \rightarrow \infty}{\cal F}(\vec{n},{\bar x})=0 \quad
\lim_{n_y\rightarrow \infty}{\cal F}(\vec{n},{\bar x})=0 \end{eqnarray} Indeed
in Eq.  (\ref{f1}), there are integrals of the form \begin{eqnarray} \int
dk_i~\psi_{ n_i}(k_i)~\psi_{0}(k_i) g_i(k_i) \quad i=x,y \end{eqnarray} where it
is easy to show that $\psi_{0}(k_i) g_i(k_i)$ is i) continuous, ii) bounded and
iii) $\in {\cal{L}}_1$ (due to the gaussian form of $\psi_{0}(k_i)$ and the
power-law behaviour of $g_i(k_i)$ for $k_i \rightarrow \pm \infty$).
Furthermore also the derivative of $\psi_{0}(k_i) g_i(k_i)$ belongs to $
{\cal{L}}_1$.  Therefore, we can exploit the same arguments presented in
Appendix C and the boundedness of ${\cal F}(\vec{n},{\bar x})$ (see Eq.
(\ref{f3})) to show that for large $n_x$ and $n_y$ one has \begin{eqnarray} &&
{\cal F}(\vec{n},{\bar x}) \leq {C \over n_x^{3/2}n_y^{3/2}} \label{lim}
\end{eqnarray}

\noindent where $C > 0$ does not depend upon $n_z$.  Actually, using the
property of the Hermite polynomials $H'_{n+1}(t) =2(n+1) H_{n}(t)$ and the
gaussian behaviour of $\psi_{0}(k_i) g_i(k_i)$, it is possible to show that the
fall-off of ${\cal F}(\vec{n},{\bar x})$ is faster than any power of $n_x n_y$.

By using the upper bound of Eq.  (\ref{lim}) one can show that even for $Q
\rightarrow \infty$ (i.e.  $n_{1(2)} \rightarrow \infty$) the sum over $n_x$ in
Eq.  (\ref{avF0}) contains only a finite number of relevant terms.  As a matter
of fact, one has \be && {1 \over (n_1-n_2)}\sum_{n_x=N_x+1}^{n_1}
\sum_{n_y=0}^{n_1-n_x}\sum_{n_z= MAX(0,n_2-n_x-n_y)}^{n_1-n_x-n_y}{\cal
F}(\vec{n},{\bar x}) \leq \nonumber \\ && {1 \over
(n_1-n_2)}\sum_{n_x=N_x+1}^{n_1} \sum_{n_y=0}^{n_1-n_x}\sum_{n_z=
MAX(0,n_2-n_x-n_y)}^{n_1-n_x-n_y} {C \over n_x^{3 / 2} n_y^{3 / 2} }\leq
\nonumber \\ && \sum_{n_x=N_x+1}^{n_1} \sum_{n_y=0}^{n_1-n_x} {C \over n_x^{3 /
2} n_y^{3 / 2}} \leq C \sum_{n_x=N_x+1}^{n_1} {1 \over n_x^{3 /
2}}\sum_{n_y=0}^{\infty} {1 \over n_y^{3 / 2}} \leq \epsilon \label{lim1} \ee

\noindent provided that $N_x$ is large enough.  By analogous arguments one can
show the same property for the sum over $n_y$.  Furthermore, the indices
$n_1-n_x-n_y$ and $n_2-n_x-n_y$ in the sum over $n_z$ can be replaced by $n_1$
and $n_2$, respectively, by using once more the upper limit in Eq.  (\ref{lim}).
As a matter of fact, for large values of $n_1$ and $n_2$ and fixed values of
$N_x$ and $N_y$, one has \be && {1 \over (n_1-n_2)}\sum_{n_x=0}^{N_x}
\sum_{n_y=0}^{N_y}~~\sum_{n_z= n_{2(1)}-n_x-n_y}^{n_{2(1)}}{\cal
F}(\vec{n},{\bar x}) \leq \nonumber \\ && {1 \over (n_1-n_2)}\sum_{n_x=0}^{N_x}
\sum_{n_y=0}^{N_y}~~\sum_{n_z= n_{2(1)}-n_x-n_y}^{n_{2(1)}} {C \over n_x^{3 / 2}
n_y^{3 / 2} }\leq \nonumber \\ && {C \over (n_1-n_2)}\sum_{n_x=0}^{N_x}
\sum_{n_y=0}^{N_y} {(n_x+n_y) \over n_x^{3 / 2} n_y^{3 / 2} }\leq \epsilon.
\label{lim2} \ee

Therefore, in the Bjorken limit, namely for large $n_1$ and $n_2$, Eq.
(\ref{avF0}) becomes \be &&{\bar F}_1({\bar x},Q)=\frac{{\bar x}^2} {16 a^2
(1-{\bar x})(n_1-n_2)}\sum_{n_x=0}^{N_x} \sum_{n_y=0}^{N_y}\sum_{n_z=
n_2}^{n_1}{\cal F}(\vec{n},{\bar x}) \label{avF1}\ee

\noindent with finite values $N_x$ and $N_y$ which do not depend upon $Q$.
 Then, adding to Eq.  (\ref{avF1}) the following sum (that is vanishing for
 large values of $N_x$ and $N_y$ due to Eq.  (\ref{lim})) \be &&{\bar S}_1({\bar
 x},Q)=\frac{{\bar x}^2} {16 a^2 (1-{\bar x})(n_1-n_2)} \left
 [\sum_{n_x=N_x+1}^{\infty} \sum_{n_y=0}^{N_y}+\sum_{n_x=0}^{\infty}
 \sum_{n_y=N_y+1}^{\infty} \right ] \cdot \nonumber \\ && \sum_{n_z=
 n_2}^{n_1}{\cal F}(\vec{n},{\bar x}) \label{avF2}\ee

\noindent one has \be &&{\bar F}_1({\bar x},Q)=\frac{{\bar x}^2} {16 a^2
(1-{\bar x})(n_1-n_2)}\sum_{n_x=0}^{\infty} \sum_{n_y=0}^{\infty}\sum_{n_z=
n_2}^{n_1}{\cal F}(\vec{n},{\bar x}) = {\rho(\bar{x}) \over 2} \ee

\noindent where Eq.  (\ref{f3}) has been used in the last step.

Arguments analogous to those presented in this subsection can be applied in
order to show that $W^{++}$ is vanishing in the Bjorken limit also in the
general case.

\section{Conclusion} \label{S5}

\begin{sloppypar} \indent Within front-front dynamics we have studied DIS in the
 Bjorken limit for a relativistic model system composed by two spin 1/2
 particles interacting via the relativistic analog of the harmonic oscillator
 potential (see Sec.  \ref{S2}).  \end{sloppypar}

 First of all, we have explicitly calculated the DIS structure functions at $Q^2
\rightarrow \infty$ for our model system with the standard parton model
assumptions in a Breit frame where the total transverse momentum ${\vec {P'}}_{
\bot}$, as well as the transverse momentum transfer ${\vec q}_{\bot}$ are zero
(i.e., in an infinite momentum frame).  We have adopted as usual a one-body
electromagnetic current, and have shown that the structure functions are indeed
given by the standard formulas of the parton model.  It should be pointed out
that a one-body choice for the current operator in the Breit reference frame is
compatible with Poincar\'e invariance (see Ref.  \cite{LPS}).

 Then we have introduced the {\em exact final state wave functions} which
properly take into account the FSI, and we have calculated the structure
functions in the same Breit frame.  In this frame, the components $J^+,J^1,J^2$
of the current operator have been taken in the one-body form, while $J^-$ has
been defined through the current conservation.  As shown in Ref.  \cite{LPS},
this choice of the current is compatible with Poincar\'e invariance and
trivially fulfills the current conservation.  We have shown that if one takes
average values of the exact structure functions over small, finite intervals
$[x_1,x_2]$ of the Bjorken variable $x$, so that $(x_2-x_1)/x_1 \ll 1$, then at
high $Q^2$ there exist many values of $n$ such that $Q^2/(Q^2+8a^2n) \in
[x_1,x_2]$ and in the Bjorken limit such average values coincide with those
given by the parton model.  This averaging procedure corresponds to the finite
resolution of the experimental measurements and allows to avoid some
mathematical technicalities of distributions.

It is the first time that the effects of the final state interactions in DIS
have been exactly calculated in a relativistic model.  It is worth noting that
the relativistic calculation differs from the nonrelativistic ones considered in
Ref.  \cite{PSW} in several aspects.  In particular, the Bjorken limit implies
that one gets a finite contribution to the structure functions only from excited
states with $n\rightarrow \infty$, while the nonrelativistic approach is valid
only if $n\ll (m_0/a)^2$ (see Eq.  (\ref {16})).  Another crucial difference
between the two cases is that the nonrelativistic relation $k_z"=k_z'+q_z/2$
considerably differs from the corresponding relativistic expression that can be
obtained by Eq.  (\ref{33}).  Furthermore in the infinite momentum frame only
the transverse components of the hadronic tensor survive in the Bjorken limit
(see Sec.  \ref{S3}), while in the nonrelativistic case the component $W^{00}$
is the dominant one.

 One might think that the main result of this paper is a consequence of the fact
that the states with given quantum numbers $(n_x,n_y,n_z)$, as well as the
states with given values of ${\vec k}$, form complete sets in ${\cal H}_{int}$
if the spin variables are dropped.  Since the hadronic tensor is determined by a
sum over all possible final states, the sum over $(n_x,n_y,n_z)$ in Eq.
(\ref{37}) should give the same result as the integration over ${\vec k}$ in Eq.
(\ref{20}), as a consequence of the above mentioned completeness.  This is not
the case due to the presence of the delta function in Eq.  (\ref{37}).  Indeed,
at fixed values of $x$ and $Q$, the sum over $(n_x,n_y,n_z)$ is carried out only
at some fixed values of the three indices (see Eq.  (\ref{40})).  Then the
calculation of the average values of the structure functions (see Eqs.
(\ref{50}) and (\ref{52})) involves only a small part of all values of $n \equiv
n_z$.  Therefore formally the completeness property cannot be used.
Heuristically, one can say that the eigenstates of the relativistic analog of
the harmonic oscillator potential are equivalent to the free states in the
relevant part of the Hilbert space.  It represents an interesting topic to be
investigated whether the same property holds for different confining
interactions.  An analogous result on the equivalence between interacting and
free eigenstates has been obtained for the reaction $e^+e^- \rightarrow hadrons$
in the model considered in \cite{sim}.

 Our results could be considered as an argument in favor of the "common wisdom",
according to which the FSI in the Bjorken limit is a higher twist effect (in
this connection it could be interesting to calculate the terms of order $1/Q^2$
in the structure functions and to see how they depend on the confinement
radius).  As noted above, our choice of the current is a possible choice,
compatible with Poincar\'e invariance and current conservation.  However, these
requirements do not determine the current operator uniquely and many body
components could be present in $J^{\mu}(0)$.  Therefore one should study whether
the results of the parton model can still be recovered in the Bjorken limit if
the operator $J^{\mu}(0)$ contains many-body interaction terms.  This problem
will be considered elsewhere within a more refined model, including possibly a
three particle system.

\newpage \begin{center} {\large \bf Acknowledgments} \end{center}

The authors are grateful to S.J.  Brodsky, A.B.  Kaidalov, B.Z.  Kopeliovich,
L.N.  Lipatov and H.C.  Pauli for useful discussions.  The work of one of the
authors (FML) was supported in part by grant No.  96-02-16126a from the Russian
Foundation for Fundamental Research.

\begin{center} {\large\bf Appendix A} \end{center}

\setcounter{equation}{0} \def\theequation{A.\arabic{equation}}

\begin{sloppypar} In this Appendix it will be proved that \begin{equation}
\lim_{n\rightarrow \infty} f(n,x) = \frac{(8\pi)^{1/2}a\chi_0(x)}{m_0x^{3/2}}
\nonumber \end{equation} \noindent From Eqs.  (\ref{15}) and (\ref{49}) one has
\begin{equation} f(n,x)=f_1(n,x)+f_2(n,x)+f_3(n,x)+f_4(n,x)+f_5(n,x) \label{56}
\end{equation}

\noindent where \begin{equation} f_1(n,x)=\frac{a\chi_0(x)}{m_0x^{3/2}}
(8\sqrt{\pi})^{1/2}f(n), \label{57} \end{equation} \begin{eqnarray}
&&f_2(n,x)=(-1)^n(2\sqrt{\pi})^{1/2} \int_{-\infty}^{-b_1} \Phi_n(t) \cdot
\nonumber\\ && [\frac{\chi_0(\xi')}{\xi'(\xi'\xi)^{1/2}
(t^2+m_0^2/a^2)^{1/4}}-\frac{2a\chi_0(x)|t|^{1/2}}{m_0x^{3/2}}]dt, \label{59}
\end{eqnarray}

\begin{eqnarray} &&f_3(n,x)=(-1)^n(2\sqrt{\pi})^{1/2} \int_{-b_1}^{0} \Phi_n(t)
\cdot \nonumber\\ && [\frac{\chi_0(\xi')}{\xi'(\xi'\xi)^{1/2}
(t^2+m_0^2/a^2)^{1/4}}-\frac{2a\chi_0(x)|t|^{1/2}}{m_0x^{3/2}}]dt, \label{61}
\end{eqnarray}

\begin{equation} f_4(n,x)=(-1)^n(2\sqrt{\pi})^{1/2}\int_{0}^{b_2} \Phi_n(t)
\frac{\chi_0(\xi')}{\xi'(\xi'\xi)^{1/2} (t^2+m_0^2/a^2)^{1/4}}dt, \label{62}
\end{equation}

\begin{equation} f_5(n,x)=(-1)^n(2\sqrt{\pi})^{1/2}\int_{b_2}^{\infty} \Phi_n(t)
\frac{\chi_0(\xi')}{\xi'(\xi'\xi)^{1/2} (t^2+m_0^2/a^2)^{1/4}}dt, \label{63}
\end{equation}

\noindent with $f(n)$ given by

\begin{equation} f(n)=\int_{0}^{\infty} \Phi_n(t)t^{1/2} dt, \label{58}
\end{equation}

\noindent and $\Phi_n(t)$ defined in Eq.  (\ref{15}).

 So far no approximation is made and $b_1$ and $b_2$ are arbitrary positive
numbers.

 In order to prove Eq.  (\ref{55}), we have to show that for any $\epsilon > 0$
there exists $N$ such that

\begin{equation} |f(n,x)-\frac{(8\pi)^{1/2}a\chi_0(x)}{m_0x^{3/2}}|<\epsilon
\quad for\, all \quad n>N \label{64} \end{equation}

 Let us first consider $f_5(n,x)$.  When $t=k_z/a$ is positive and large then,
as follows from Eq.  (\ref{45}), $\xi$ is close to 1.  In turn, as follows from
Eq.  (\ref{42}), $\xi'-\xi$ is small.  Therefore for large $t$ we can replace
$\xi'$ by $\xi$ and then $\chi_0(\xi')$ by $\chi_0(\xi)$, which falls off
exponentially when $t\rightarrow \infty$.  Let us use the property (see Sec.
10.18 of Ref.  \cite{BE})

\begin{equation} |\Phi_n(t)| < K \label{65} \end{equation}

\noindent where $K\approx 1.086435$.  Therefore, for a given value of $\epsilon$
it is possible to find $b_2$ such that $|f_5(n,x)|<\epsilon /5$ for all $n$.
Due to Eq.  (\ref{65}), the quantity $b_2$ depends only on $\epsilon$, but not
on $n$.

  As far as $f_4(n,x)$ is regarded, we note that, as follows from Eqs.
(\ref{15}), (\ref{42}) and (\ref{45}), there exist positive quantities $c_1(x)$
and $c_2(x)$ such that

\begin{equation} c_1(x)\leq
\frac{\chi_0(\xi')}{\xi'(\xi'\xi)^{1/2}(m_0^2+k_z^2)^{1/4}} \leq c_2(x)
\label{66} \end{equation}

\noindent when $t\in [0,b_2]$.  Now we use the fact \cite{BE} that, uniformly in
any bounded interval,

\begin{equation} \Phi_n(t) = \frac{1}{(2^nn!)^{1/2}}
\frac{\Gamma(n+1)}{\Gamma(n/2+1)} [cos(N^{1/2}t-\frac{n\pi}{2})+O(n^{-1/2})]
\label{67} \end{equation}

\noindent where $N=2n+1$.  Therefore, using the Stirling formula \cite{BE}, for
large $n$ one has

\begin{equation} \Phi_n(t) = (\frac{2}{\pi n})^{1/4} \left[ cos(N^{1/2}t-n\pi
/2) + O(n^{-1/2}) \right].  \label{68} \end{equation}

 Then by the Riemann-Lebesgue theorem (see, e.g., Ref.  \cite{Titm}) it is
possible to find $N_1$ such that $|f_4(n,x)|<\epsilon /5$ for all $n>N_1$.

Let us now consider $f_2(n,x)$.  When $t$ is negative and $|t|$ is large, the
quantity $\xi$ is small and $\xi^{-1/2}$ becomes $2a|t|/m_o$ (see Eq.
(\ref{45})).  Then, as follows from Eq.  (\ref{42}), $\xi'-x$ is small and for
$t$ negative and $|t|$ large, one has

$$\frac{\chi_0(\xi')}{\xi'(\xi'\xi)^{1/2}(t^2+m_0^2/a^2)^{1/4}} \rightarrow
\frac{2a\chi_0(x)|t|^{1/2}}{m_0x^{3/2}}$$.

 An expansion of the difference between the above quantities as a Taylor series
of $1/t^2$, shows that this difference behaves as $1/|t|^{3/2}$ for large
negative values of t.  Therefore, using again Eq.  (\ref{65}), we conclude that
it is possible to find $b_1$ such that $|f_2(n,x)|<\epsilon /5$ for all $n$.
The quantity $b_1$ depends only on $\epsilon$, but not on $n$.

By analogy with the previous considerations we did for $f_4(n,x)$, it is
possible to find $N_2$, such that $|f_3(n,x)|<\epsilon /5$ for all $n>N_2$.

Finally, consider $f_1(n,x)$.  Using Eq.  (7.376) in Ref.  \cite{GR} and the
Stirling formula, for large, odd values of $n$ one has

\begin{equation} f(n) = 2 (\frac{n}{\pi})^{1/4} (-1)^{(n-1)/2} \Gamma (5/4)
F(-(n-1)/2,5/4,3/2,2) \label{69} \end{equation}

\noindent where $F(a,b,c,z)$ is the hypergeometric function.  In the literature
(see, e.g., \cite{BE}) the asymptotic behaviour of $F(a,b,c,z)$ for
$|a|\rightarrow \infty$ is usually given in the case $|z| < 1$.  In Appendix B
we obtain the asymptotic behaviour of $F(-m,b,c,z)$ for $c > b > 0$ and for
$z=2$.  Then, using this asymptotic behaviour, one finds $f(n)\rightarrow
\pi^{1/4}$ when $n\rightarrow \infty$.

 For even values of $n$ it is not convenient to use directly Eq.  (7.376) of
Ref.  \cite{GR}, since it involves $F(a,b,c,z)$ with $b > c$.  However, from the
relation \cite{BE}

\begin{equation} H_n(t) = \frac{1}{2t} \left[ H_{n+1}(t) + 2n H_{n-1}(t)\right]
\label{70} \end{equation}

\noindent one has

\begin{equation} f(n) = \frac{1}{(2^n n!)^{1/2}}\int_{0}^{\infty}
exp(-\frac{t^2}{2}) t^{-1/2} \left[ \frac{1}{2}H_{n+1}(t) + n H_{n-1}(t) \right]
dt \label{71} \end{equation}

\noindent Using again Eq.  (7.376) of Ref.  \cite{GR}, for odd values of the
Hermite polynomial index, and the Stirling formula Eq.  (\ref{71}) can be cast
in the following form for large, even values of $n$

\begin{eqnarray} f(n)& =& \frac{\Gamma (3/4) (-1)^{n/2}}{(\pi n)^{1/4}} [ (n +
1) F(-n/2,3/4;3/2;2)\nonumber\\ && - n F(-n/2 + 1,3/4;3/2;2)] \label{72}
\end{eqnarray}

Then by using the asymptotic expression for $F(-m,b,c,2)$ given in Appendix B,
also for even values of $n$ one finds $f(n)\rightarrow \pi^{1/4}$ when
$n\rightarrow \infty$.

Therefore it is possible to find $N_3$ such that

\begin{equation} |f_1(n,x)-(8\pi)^{1/2}\frac{a\chi_0(x)}{m_0x^{3/2}}
|<\epsilon/5 \quad for\, all \quad n>N_3 \label{73} \end{equation}

In conclusion, if $N=max\{N_1,N_2,N_3\}$, the condition (\ref{64}) is satisfied
and then Eq.  (\ref{55}) is proved.  \end{sloppypar}

\begin{center} {\large\bf Appendix B} \end{center}

\setcounter{equation}{0} \def\theequation{B.\arabic{equation}}

\begin{sloppypar} In this Appendix we will investigate the asymptotic behavior
of $F(-m,b,c,2)$ for $m\rightarrow \infty$.  In order to obtain this behavior
let us use the well-known fact (see, e.g., Eqs.  (9.111) and (8.834) of
\cite{GR}) that, if $c > b > 0$, then

\begin{equation} F(-m,b,c,z)=\frac{\Gamma (c)}{\Gamma (b) \Gamma
(c-b)}\int_{0}^{1} t^{b-1} (1-t)^{c-b-1} (1-tz)^{m} dt \label{74} \end{equation}

For $z = 2$, due to the factor $(1 - 2t)^{m}$ only neighborhoods of $t = 0$ and
$t = 1$ contribute to the integral when $m\rightarrow \infty$.  We can replace
$(1-t)^{c-b-1}$ by $1$ and $(1-2t)^{m}$ by $(1-t)^{2m}$ in the first
neighborhood, and $t^{b-1}$ by 1 and $(1-2t)^{m}$ by $(-1)^m t^{2m}$ in the
second one.  Then Eq.  (\ref{74}) becomes

\begin{eqnarray} F(-m,b,c,2)=\frac{\Gamma (c)}{\Gamma (b) \Gamma (c-b)} \left[
\int_{0}^{1} t^{b-1} (1-t)^{2m} dt ~~+~ \right .  \nonumber\\ \left .
+~~\int_{0}^{1} (1-t)^{c-b-1} (-1)^m t^{2m} dt \right] , \label{75}
\end{eqnarray}

\noindent since in the first integral there is a non-zero contribution only in
the neighborhood of $t = 0$ and in the second integral only in the neighborhood
of $t = 1$.  Then, using the definition of the function $B(x,y)$ (see, e.g., Eq.
(8.380) of \cite{GR}) and its expression in terms of the function $\Gamma (z)$
(see, e.g., Eq.  (8.384) of \cite{GR}) one obtains

\begin{eqnarray} F(-m,b,c,2) = \frac{\Gamma (c) \Gamma (2m + 1)}{\Gamma (c-b)
\Gamma (b + 2m + 1)} ~+ \nonumber\\ +~~(-1)^m \frac{\Gamma (c) \Gamma (2m +
1)}{\Gamma (b) \Gamma (c-b + 2m + 1)} \label{76} \end{eqnarray}

Finally, using the Stirling formula, one finds the following asymptotic
behaviour for $m\rightarrow \infty$

\begin{equation} F(-m,b,c,2) = \frac{\Gamma (c)}{\Gamma (c-b) (2m + 1)^b} +
(-1)^m \frac{\Gamma (c)}{\Gamma (b) (2m + 1)^{(c-b)}}.  \label{77}
\end{equation} \end{sloppypar}

\begin{center} {\large\bf Appendix C} \end{center}

\setcounter{equation}{0} \def\theequation{C.\arabic{equation}}

\begin{sloppypar} In this Appendix it will be proved that $W^{++}$ is vanishing
in the Bjorken limit.

Since from Eqs.  (\ref{30}) and (\ref{Breit}) one has $p^{"+}p^{'+}=\xi'\xi
Q(1-x)/x$ and from the delta function on the masses (see, e.g., Eq.  (\ref{48}))
one obtains $Q=2 a \sqrt{2n} \sqrt{x/(1-x)}$, then apart from constant factors
the relevant integral in the calculation of $W^{++}$ is \be I_n=\sqrt{n}
\int^\infty_{-\infty} \Phi_n(t) \sqrt{\xi \over \xi'} {\chi_0(\xi') \over
(t^2+m_0^2/a^2)^{1/4}}dt = \nonumber \\\sqrt{n} ~\int^\infty_{-\infty} \Phi_n(t)
\Psi(t)dt = \sqrt{n}~ \lim_{\stackrel{\scriptstyle{b\rightarrow
\infty}}{c\rightarrow \infty}}~\int^b_{-c} \Phi_n(t) \Psi(t)dt \ee with
$\Psi(t)= \sqrt{(\xi / \xi')}~\chi_0(\xi') / (t^2+m_0^2/a^2)^{1/4}$.  Let us
investigate the limit \be \lim_{n\rightarrow \infty}\sqrt{n} \int^b_{-c}
\Phi_n(t) \Psi(t)dt \label{b3} \ee We will show that this limit is uniform with
respect to the extrema.  First of all, let us observe that the function
$\Psi(t)$ is i) continuos, ii) bounded and iii) $\in {\cal{L}}_1$.  Indeed, it
falls exponentially for ${t\rightarrow \infty}$ and as $|t|^{-3/2}$ for
${t\rightarrow~ -\infty}$ (as it can be argued from the discussion in Appendix
A).  Furthermore also the derivative of $\Psi(t)$ $\in {\cal{L}}_1$, due to the
behavior of $\Psi(t)$ for $ t \rightarrow \pm \infty$.  Since we have a bounded
interval in (\ref{b3}), we can use the asymptotic expression (\ref{68})) for
$\Phi_n(t)$ \be \Phi_n(t) = (\frac{2}{\pi n})^{1/4}
[cos(N^{1/2}t-\frac{n\pi}{2})+O(n^{-1/2})] \label{b4} \ee with $N=2n+1$.  The
second term in Eq.  (\ref{b4}) gives an integral in Eq.  (\ref{b3}) that
uniformly vanishes as $1/n^{1/4}$ with respect to the integration extrema, due
to the property iii).  Also the first term in Eq.  (\ref{b4}) produces an
integral in Eq.  (\ref{b3}) that uniformly vanishes as $1/n^{1/4}$ with respect
to the integration extrema.  This can be shown with an integration by parts and
exploiting the property ii) and the integrability of $|d\Psi(t)/dt|$.

Therefore the limit in Eq.  (\ref{b3}) is zero and it is uniform with respect to
the integration extrema.  Then one has \be &
&\lim_{\stackrel{\scriptstyle{b\rightarrow \infty}}{c\rightarrow
\infty}}~\lim_{n\rightarrow \infty}\sqrt{n}\int^b_{-c} \Phi_n(t) \Psi(t)dt =
\nonumber \\ & &\lim_{n\rightarrow
\infty}~\lim_{\stackrel{\scriptstyle{b\rightarrow \infty}}{c\rightarrow
\infty}}\sqrt{n}\int^b_{-c} \Phi_n(t) \Psi(t)dt \ee and finally \be
&&\lim_{n\rightarrow \infty} I_n =0 \ \ee \end{sloppypar}

\end{document}